\def\spose#1{\hbox to 0pt{#1\hss}}
\def\lta{\mathrel{\spose{\lower 3pt\hbox{$\mathchar"218$}}
     \raise 2.0pt\hbox{$\mathchar"13C$}}}
\def\gta{\mathrel{\spose{\lower 3pt\hbox{$\mathchar"218$}}
     \raise 2.0pt\hbox{$\mathchar"13E$}}}
\def\ge{\mathrel{\spose{\lower 3pt\hbox{$-$}}
     \raise 2.0pt\hbox{$\mathchar"13E$}}}
\def\le{\mathrel{\spose{\lower 3pt\hbox{$-$}}
     \raise 2.0pt\hbox{$\mathchar"13C$}}}

\documentclass[aps,twocolumn]{revtex4}

\usepackage{graphics}

\begin{document}

\bibliographystyle{apsrev}

\title{Detecting a stochastic gravitational wave background with
the Laser Interferometer Space Antenna}
\author{Neil J. Cornish}
\affiliation{Department of Physics, Montana State University, Bozeman, MT 59717}

\begin{abstract}
The random superposition of many weak sources will produce a stochastic
background of gravitational waves that may
dominate the response of the LISA (Laser Interferometer Space Antenna)
gravitational wave observatory. Unless something can be done to distinguish
between a stochastic background and detector noise, the two will combine to
form an effective noise floor for the detector. Two methods have been proposed
to solve this problem.
The first is to cross-correlate the output of two independent interferometers.
The second is an ingenious scheme for monitoring the instrument noise by operating
LISA as a Sagnac interferometer.
Here we derive the optimal orbital alignment for cross-correlating
a pair of LISA detectors, and provide the first analytic derivation of the Sagnac
sensitivity curve.
\end{abstract}

\maketitle


\section{Introduction}

It is hoped that the Laser Interferometer Space Antenna (LISA)\cite{lppa}
will be in operation by 2011. To meet this
deadline, basic design decisions need to be made in the next few
years. One decisions concerns the gravitational wave background.
Depending on ones point of view, the gravitational wave background is either
a blessing or a curse. Those hoping to use LISA to
observe black hole coallesence see the stochastic
background as a potential source of noise, while those hoping to use LISA
to study binary populations see the stochastic background as a promising
source of information. But for the gravitational wave background to be of any use,
a way has to be found to distinguish it from instrument noise.

One would have to have great faith in the theoretical noise model to claim that
excess noise in the LISA detector was due to a stochastic background of gravitational waves.
However, with two independent Michelson interferometers\cite{mich,paper1}, or a combined
Michelson-Sagnac interferometer\cite{aet,hogb}, there are ways to separate the signal from
the noise.
We will review both of these approaches and derive several new results relating to
each method. Our main result is a derivation of the optimal orbital alignment to use
when cross-correlating two LISA detectors.

The outline of the paper is as follows. In Section II we derive the response of
Michelson and Sagnac interferometers to a plane, monochromatic gravitational wave.
In Section III the detector responses are used to derive sensitivity curves for
the interferometers responding to a stochastic background of gravitational waves.
Section IV discusses the cross-correlation of two detectors. Section V is devoted to
optimizing the cross-correlation of two LISA detectors. In Section VI, the results of
Sections II through V are applied to the problem of detecting a stochastic background
of gravitational waves from White Dwarf binaries and Inflation.

\section{Detector Response}

The proper distance between two freely moving masses fluctuates when a
gravitational wave passes between them. Suppose that ${\bf r}$ is a unit
vector pointing from Mass 1 to Mass 2, and $L$ is the proper distance between
the masses in the absence of gravitational waves. Together these masses can
form one arm of a gravitational wave interferometer. Now suppose that a plane
gravitational wave, described in the transverse-traceless gauge by the
tensor ${\bf h}(f,t,{\bf x})$, propagates in
the $\widehat{\Omega}$ direction with frequency $f$. A photon leaving
Mass 1 (located at ${\bf x}_1$)
at time $t_1$ will travel a proper distance
\begin{equation}\label{onearm}
\ell_{12}(t_1) = L\left(1 + {\bf h}(f,t_1,{\bf x}_1):{\bf D}(\widehat\Omega,f)\right)\, ,
\end{equation}
to reach Mass 2. Here
\begin{equation}
{\bf D}(\widehat\Omega,f)=\frac{1}{2}({\bf r}\otimes{\bf r})
{\cal T}({\bf r}\cdot\widehat{\Omega},f)
\end{equation}
is the detector tensor for the arm and
\begin{equation}
{\cal T}({\bf r}\cdot\widehat{\Omega},f)={\rm sinc}\left[\frac{f}{2 f_*}
\left(1-{\bf r}\cdot \widehat{\Omega}\right)\right]e^{i\frac{f}{2 f_*}
\left(1-{\bf r}\cdot \widehat{\Omega}\right)}
\end{equation}
is the transfer function. The characteristic frequency scale of the detector
is given by $f_*=c/(2\pi L)$.

\begin{figure}[ht]
\vspace{60mm}
\includegraphics{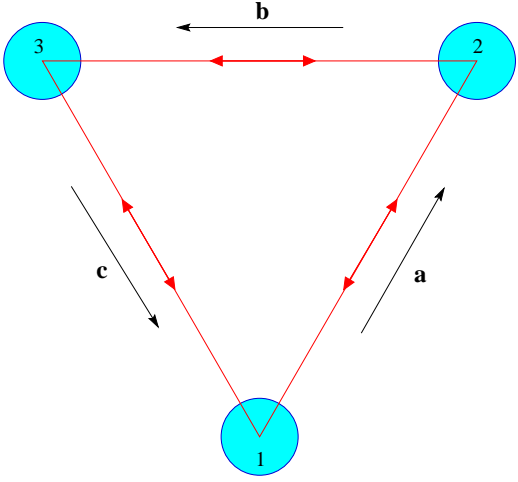}
\vspace{8mm}
\caption{Laser signals used to track the LISA constellation.}
\end{figure} 

With perfectly stable lasers it is possible to build a one-arm gravitational wave detector.
The phase of light making a round trip down the arm can be compared to the phase of
light stored in the laser cavity. The phase shift measures the
change in proper distance along the arm. However, laser phase noise
prevents us from building a viable one-arm interferometer.
The simplest way to eliminate laser phase noise is to compare signals that have
traveled approximately the same distance. This is the approach taken in LISA
Pre-Phase A Report\cite{lppa}, where it is proposed that three masses be placed at
the vertices of an equilateral triangle, and the phase shift in the round-trip laser
signal along two of the arms be used to monitor changes in the proper distance between
the masses. In other words, the plan is to build a space-based Michelson interferometer.
Referring to the diagram in Figure 1, we see that there are three ways of forming a
Michelson interferometer from the LISA triangle. The result (\ref{onearm})
for the variation in the length of a single
arm can be used to derive the response of the Michelson interferometers.
For example, the interferometer with vertex ${\bf x}_1$ experiences a phase
variation of
\begin{eqnarray}\label{m1}
s_1(t) &=& \frac{1}{2L}\big(\ell_{12}(t-2L)+\ell_{21}(t-L)  \nonumber \\
&& \quad -\ell_{13}(t-2L)-\ell_{31}(t-L)\big) \nonumber \\
&& \nonumber \\
&=& {\bf D}_m(\widehat\Omega,f):{\bf h}(f,t,{\bf x}_1)\, ,
\end{eqnarray}
where
\begin{eqnarray}
{\bf D}_{\rm m}(\widehat\Omega,f) &=& \frac{1}{2}\left( ({\bf a}\otimes{\bf a})
        \, {\cal T}_{\rm m}({\bf a}\cdot\widehat{\Omega},f) \right. \nonumber \\
       && - \left.
    ({\bf c}\otimes{\bf c})\, {\cal T}_{\rm m}(-{\bf
     c}\cdot\widehat{\Omega},f) \right)
\end{eqnarray}
and
\begin{eqnarray}
    && {\cal T}_{\rm m}({\bf u}\cdot\widehat{\Omega},f)= \nonumber \\
    && \; \frac{1}{2}\left[
    {\rm sinc}\left( \frac{f(1-{\bf
    u}\cdot\widehat{\Omega})}{2f_* }\right)\exp\left(-i\frac{f}{2f_*}(3 +{\bf
    u}\cdot\widehat{\Omega})\right) \right. \nonumber \\
    &&\; + \left. {\rm sinc}\left(\frac{f(1+{\bf
    u}\cdot\widehat{\Omega})}{2f_* }\right) \exp\left(-i\frac{f}{2f_*}(1+{\bf
    u}\cdot\widehat{\Omega})\right)\right] 
\end{eqnarray}

There are many other ways to combine the lasers signals in the LISA triangle.
A particularly useful combination comes from comparing the phase of signals that
are sent clockwise and counter-clockwise around the triangle. An interferometer of
this type was built by Sagnac\cite{sag} to study rotating frame effects.
The Sagnac signal extracted at vertex 1 is given by
\begin{eqnarray}
s_1(t)&=&\frac{1}{3L}\big( \ell_{13}(t-3L)+\ell_{32}(t-2L)+\ell_{21}(t-L)
\nonumber \\
&& - \ell_{12}(t-3L)-\ell_{23}(t-2L)-\ell_{31}(t-L) \big) \nonumber \\
&=& {\bf D}_{\rm s}(\widehat\Omega,f):{\bf h}(f,t,{\bf x}_1) \, ,
\end{eqnarray}
where
\begin{eqnarray}
{\bf D}_{\rm s}(\widehat\Omega,f) &=& \frac{1}{6}\left( ({\bf a}\otimes{\bf a})
        \, {\cal T}_a(f) +
      ({\bf b}\otimes{\bf b})\, {\cal T}_b(f)\right. \nonumber \\
       && + \left.
    ({\bf c}\otimes{\bf c})\, {\cal T}_c(f) \right)
\end{eqnarray}
and
\begin{eqnarray}
{\cal T}_a(f)&=&
e^{-i\frac{f}{f_*}(1+{\bf a}\cdot\widehat{\Omega})}{\rm sinc}\left(\frac{f}{2f_*}\left(
1+{\bf a}\cdot\widehat{\Omega}\right)\right) \nonumber \\
&& - e^{-i\frac{f}{f_*}(5+{\bf a}\cdot\widehat{\Omega})}{\rm sinc}\left(\frac{f}{2f_*}\left(
1-{\bf a}\cdot\widehat{\Omega}\right)\right) \nonumber \\  \nonumber \\
{\cal T}_b(f)&=& e^{-i\frac{f}{f_*}(3+({\bf a}-{\bf c})\cdot\widehat{\Omega})}\left(
{\rm sinc}\left(\frac{f}{2f_*}\left(1+{\bf b}\cdot\widehat{\Omega}\right)\right) \right.
\nonumber \\
&& \left. \quad -{\rm sinc}\left(\frac{f}{2f_*}\left(1-{\bf b}\cdot\widehat{\Omega}\right)
\right)\right)\nonumber \\ \nonumber \\
{\cal T}_c(f) &=& e^{-i\frac{f}{f_*}(5-{\bf c}\cdot\widehat{\Omega})}{\rm sinc}
\left(\frac{f}{2f_*}\left(
1+{\bf c}\cdot\widehat{\Omega}\right)\right)\nonumber \\
&& - e^{-i\frac{f}{f_*}(1-{\bf c}\cdot\widehat{\Omega})}{\rm sinc}\left(\frac{f}{2f_*}\left(
1-{\bf c}\cdot\widehat{\Omega}\right)\right)
\end{eqnarray}

Even more useful than the basic Sagnac signal is the symmetrized Sagnac signal
formed by averaging the output from the three vertices:
\begin{eqnarray}
s(t)&=&\frac{1}{3}\left(s_1(t)+s_2(t)+s_3(t)\right) \nonumber \\
&=& {\bf D}_{\rm ss}(\widehat\Omega,f):{\bf h}(f,t,{\bf x}_1)
\end{eqnarray}
where
\begin{eqnarray}
{\bf D}_{\rm ss}(\widehat\Omega,f) &=& \frac{1}{6}\left( ({\bf a}\otimes{\bf a})
        \, {\cal T}_{\rm s}({\bf a}\cdot\widehat{\Omega},f) +
      ({\bf b}\otimes{\bf b})\, {\cal T}_{\rm s}({\bf b}\cdot\widehat{\Omega},f)\right. \nonumber \\
       && + \left.
    ({\bf c}\otimes{\bf c})\, {\cal T}_{\rm s}({\bf c}\cdot\widehat{\Omega},f) \right)
\end{eqnarray}
and
\begin{eqnarray}
&&{\cal T}_{\rm s}({\bf u}\cdot\widehat{\Omega},f)=\left(1+2\cos\frac{f}{f_*}\right)
e^{-i\frac{f}{2f_*}(3+{\bf u}\cdot\widehat{\Omega})} \\
&& \times \left({\rm sinc}\left(\frac{f}{2f_*}\left(1+{\bf u}\cdot\widehat{\Omega}\right)\right)
-{\rm sinc}\left(\frac{f}{2f_*}\left(1-{\bf u}\cdot\widehat{\Omega}\right)\right)\right)\, .
\nonumber
\end{eqnarray}

The magnitude of the detector tensors ${\bf D}_{\rm m}$, ${\bf D}_{\rm s}$ and ${\bf D}_{\rm ss}$
decay as $f^{-1}$ for $f\gg f_*$. At low frequencies, $f\ll f_*$, the Michelson
interferometer has a flat response,
${\bf D}_{\rm m} \sim f^0$, while the Sagnac response decays as
${\bf D}_{\rm s}\sim f$ and the symmetrized Sagnac response decays as ${\bf D}_{\rm ss}\sim f^2$.
The insensitivity of the symmetrized Sagnac interferometer to low frequency gravitational waves
makes it the perfect tool for monitoring instruments noise in the Michelson signal\cite{aet}.

\section{Sensitivity Curves}

The detector responses derived in the last section can be used to find
the sensitivity of the interferometers to a stochastic background of gravitational
waves. A stochastic background can be expanded in terms of plane waves:
\begin{eqnarray}\label{sto}
    && h_{ij}(t,{\bf x}) =
    \int_{-\infty}^{\infty} df \, \int d\widehat{\Omega} \;
    \tilde{h}_{ij}(\widehat\Omega,f,{\bf x},t) \nonumber \\
    && = \sum_{A=+,\times}
    \int_{-\infty}^{\infty} df \int d\widehat{\Omega} \;
    \tilde{h}_A(f,\widehat{\Omega})e^{2\pi i f (t -
    \widehat{\Omega}\cdot{\bf x})}
    e_{ij}^{A}(\widehat{\Omega}) .\nonumber \\
\end{eqnarray}
Here $\int d\widehat{\Omega}$ denotes an integral over the celestial sphere and
$\tilde{h}_A(-f)=\tilde{h}^*_A(f)$ are the Fourier amplitudes of the wave.
The sum is over the two polarizations of the gravitational wave with
basis tensors $e_{ij}^+$ and $e_{ij}^\times$. Each component of the
decomposition is a plane wave with frequency $f$ propagating in the
$\widehat{\Omega}$ direction. We assume that the background can be treated as a
stationary, Gaussian random process characterized by the expectation values
\begin{eqnarray}\label{gauss}
&& \langle
    \tilde{h}^*_A(f,\widehat{\Omega})
    \tilde{h}_{A'}(f',\widehat{\Omega}')\rangle
    =  \frac{1}{2}\delta(f-f')
    \frac{\delta^2(\widehat{\Omega},\widehat{\Omega}')}
    {4\pi}\delta_{AA'} \, S_h(f) \nonumber \\
&& \langle \tilde{h}_A(f,\widehat{\Omega})\rangle =  0,
\end{eqnarray}
where $S_h(f)$ is the one-sided power spectral density. The noise in the detector
is treated as a Gaussian random process with zero mean and one-sided spectral density
$S_n(f)$. The total output of the interferometer, $S(t)$, is a combination of
signal and noise: $S(t)=s(t)+n(t)$. The results from section II, in conjunction
with equations (\ref{sto}) and (\ref{gauss}), yield
$\langle  S(t)\rangle = 0$ and
\begin{eqnarray}
\langle  S^2(t)\rangle &=& 
\langle  s^2(t)\rangle +2\langle s(t)n(t) \rangle + \langle n^2(t) \rangle \nonumber \\
&=& \langle  s^2(t)\rangle + \langle n^2(t) \rangle \nonumber \\
&=& \int_0^\infty df S_h(f) {\cal R}(f) + \int_0^\infty df S_n(f) \, .
\end{eqnarray}
The interferometer response function is defined by
\begin{equation}\label{rtran}
     {\cal R}(f) = \int \frac{d \widehat{\Omega}}{4\pi} \sum_{A}
    {F^A}^*(\widehat{\Omega},f)F^A(\widehat{\Omega},f) \, ,
\end{equation}
where
\begin{equation}
F^A(\widehat{\Omega},f) = {\bf D}(\widehat\Omega,f):{\bf e}^{A}(\widehat\Omega) \, 
\end{equation}
is the antenna pattern and ${\bf D}(\widehat\Omega,f)$ is any of the detector tensors derived in section II.
The integral in (\ref{rtran}) can be done analytically in the high and low frequency
limits. The response of the Michelson, Sagnac and symmetrized
Sagnac interferometers in the low frequency limit is given by
\begin{eqnarray}
 {\cal R}_{\rm m}(f) & = & \frac{3}{10}-\frac{507}{5040}\left(\frac{f}{f_{*}}\right)^2+\dots
\nonumber \\ \nonumber \\
 {\cal R}_{\rm s}(f) & = & \frac{2}{15} \left(\frac{f}{f_{*}}\right)^2
-\frac{839}{15120}\left(\frac{f}{f_{*}}\right)^4+\dots
\nonumber \\ \nonumber \\
 {\cal R}_{\rm ss}(f) & = & \frac{1}{3024} \left(\frac{f}{f_{*}}\right)^4
-\frac{19}{72576}\left(\frac{f}{f_{*}}\right)^6+\dots \, .
\end{eqnarray}
The comparison between the Michelson and symmetrized Sagnac interferometers is
particularly striking.

The noise spectral density in the interferometer output combines all the noise contributions
along the optical path with appropriate noise transfer functions. The noise spectral density
in each signal is derived in the appendix, where it is found that
\begin{eqnarray}
S^{\rm m}_n(f)&=&4 S_s(f)+8(1+\cos^2(f/f_*))S_a(f)
\nonumber \\
&& \nonumber \\
S^{\rm s}_n(f)&=&6S_s(f)+8\left(\sin^2(3f/2f_*)+2\sin^2(f/2f_*)\right)S_a(f)\nonumber \\
&& \nonumber \\
S^{\rm ss}_n(f)&=&\frac{2}{3}\left(1+2\cos(f/f_*)\right)^2\left(S_s(f)\right. \nonumber \\
&& \left. \hspace*{1.1in}+ 4\sin^2(f/2f_*)S_a(f)\right)\, .
\end{eqnarray}
These estimates include contributions from shot noise in the photo detectors, $S_s(f)$,
and acceleration noise from the drag-free system $S_a(f)$. Using the noise budget quoted in
the LISA pre-Phase A report, we take these to equal
\begin{eqnarray}
&&S_s(f) = 4.84 \times 10^{-42} \; {\rm Hz}^{-1}\nonumber \\
&&S_a(f) = 2.31 \times 10^{-40} \left(\frac{{\rm mHz}}{f}\right)^4 \quad {\rm Hz}^{-1} \, .
\end{eqnarray}

\begin{figure}[ht]
\vspace{53mm}
\includegraphics{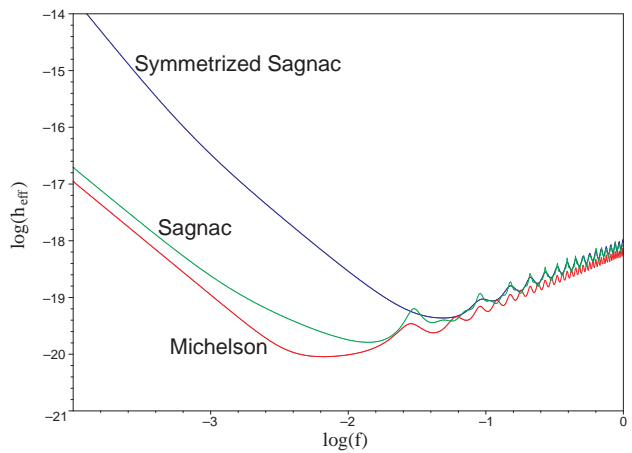}
\vspace{8mm}
\caption{Sensitivity curves for LISA operating as a Michelson, Sagnac and
symmetrized Sagnac interferometer. The frequency is measured in Hz and the strain
spectral density, $\tilde{h}_{\rm eff}(f)$, has units of Hz$^{-1/2}$.}
\end{figure}

The spectral densities $S_h(f)$ and $S_n(f)$ are related to the strain spectral densities
in the interferometer, $\tilde{h}_s(f)$ and $\tilde{h}_n(f)$:
\begin{equation}
\tilde{h}_n(f)=\sqrt{S_n(f)} \quad {\rm and} \quad \tilde{h}_s(f)=\sqrt{S_h(f){\cal R}(f) } .
\end{equation}
The integrated signal-to-noise ratio is defined:
\begin{equation}
{\rm SNR} = \frac{\langle  s^2(t)\rangle}{\langle n^2(t) \rangle}\, ,
\end{equation}
while the contribution to the SNR from a frequency band of width $\Delta f$, centered
at $f$ is given by
\begin{equation}
{\rm SNR}(f) = \frac{S_h(f) {\cal R}(f)}{S_n(f)} = 
\left(\frac{\tilde{h}_s(f)}{\tilde{h}_n(f)}\right)^2 .
\end{equation}
Sensitivity curves for space-based interferometers usually display
some multiple of the effective strain noise
\begin{equation}\label{hef}
\tilde{h}_{\rm eff}(f)=\sqrt{\frac{S_n(f)}{{\cal R}(f)}} \, .
\end{equation}
To have a signal-to-noise of one, a source of gravitational waves must have a strain
spectral density $\tilde{h}_s(f)$ that exceeds $\tilde{h}_{\rm eff}(f)$. The convention
in the LISA community is to set a signal-to-noise threshold of five (in terms of
spectral power), so standard sensitivity curves display $\sqrt{5}\tilde{h}_{\rm eff}(f)$.
However, we prefer to plot $\tilde{h}_{\rm eff}(f)$ directly. Sensitivity curves for
LISA are shown in Figure 2. The sensitivity curves all scale as $f$ in the high frequency
limit. In the low frequency limit the Michelson
and Sagnac curves scale as $f^{-2}$, while the symmetrized Sagnac sensitivity curve
scales as $f^{-3}$.
The basic Sagnac configuration is only slightly less sensitive
than the standard Michelson configuration. However, below the LISA transfer
frequency of $f_*=9.54$ mHz, the symmetrized Sagnac interferometer is considerably
less sensitive to a stochastic background than the Michelson configuration.
Unless the amplitude of the
stochastic background exceeds current predictions\cite{hb,raf} by several orders of magnitude, the
output of the symmetrized Sagnac interferometer will be all noise and no signal. Thus,
the symmetrized Sagnac signal can be used to monitor instrument noise in the
more sensitive Michelson interferometer\cite{aet,hogb}.

\section{Cross-correlating two detectors}

While monitoring the detector noise with the Sagnac signal is a great idea in theory, it
may run into problems in practice. For one, the noise in the symmetrized Sagnac interferometer
involves a slightly different combination of acceleration and position noise than is found in
the symmetrized Michelson interferometers\cite{foot}, making it an imperfect monitoring tool.
Of even
greater concern is the lack of redundancy in the Sagnac signal. If just one of LISA's
six photo-detectors fails, the Sagnac signal is lost. For these reasons we favor an
alternative strategy that works by cross-correlating the output of two fully independent
interferometers. 
The advantage of a two detector system is that while the gravitational wave signal is
correlated in each detector, the noise is not.  Thus, the signal-to-noise ratio in
the cross-correlated detector output will grow as the square
root of the observation time (for Gaussian noise). 
Similar reasoning led to the building of two rather than one
ground-based LIGO (Laser Interferometer Gravitational wave Observatory) detectors. 
The disadvantage of a two detector
observatory  is that it costs more to build, launch and operate. 
However, economy of scale suggests that the costs would not double, and
having a total of six spacecraft greatly improves the redundancy of the mission. As
many as three spacecraft could fail and still leave a working interferometer.
In contrast, the current LISA design can not afford to lose any spacecraft.

In this section we derive the sensitivity of an arbitrarily oriented pair
of interferometers to a stochastic gravitational wave background. 
We begin by considering the simple equal time correlation, $S_1(t)S_2(t)$, of the
detector outputs. The expectation value of this correlator,
\begin{eqnarray}
\langle S_1(t)S_2(t) \rangle &=&  \langle s_1(t)s_2(t) \rangle +  \langle s_1(t)n_2(t) \rangle
 \nonumber \\ &+& \langle n_1(t)s_2(t) \rangle + \langle n_1(t)n_2(t) \rangle \nonumber \\
&=& \langle s_1(t)s_2(t) \rangle \, , 
\end{eqnarray}
involves the signal in each interferometer but not the noise. Using the results of the
previous sections we find
\begin{equation}
\langle S_1(t)S_2(t) \rangle = \int_0^\infty df S_h(f) {\cal R}_{12}(f)
\end{equation}
where
\begin{equation}\label{r12}
    {\cal R}_{12}(f) = \sum_A \int \frac{d
    \widehat{\Omega}}{4\pi}
    {F_1^{A}}^*(\widehat{\Omega},f)F_2^{A}(\widehat{\Omega},f)
    e^{2\pi i f \widehat{\Omega}\cdot({\bf x}_1-{\bf x}_2)} .
\end{equation}
Here ${\bf x}_1$ and ${\bf x}_2$ are the position vectors of the corner spacecraft
in each interferometer. For coincident and coaligned detectors, ${\cal R}_{12}(f)$
approaches $2/5\sin^2\beta$ in the low frequency limit, where $\beta$ is the angle between
the interferometer arms. The overlap reduction function, $\gamma(f)$, describes how the
cross-correlation is affected by the geometry of the detector pair. The overlap reduction function
is obtained by normalizing ${\cal R}_{12}(f)$ by its low frequency limit:
\begin{equation}
\gamma(f) = \frac{5}{2\sin^2\beta} {\cal R}_{12}(f) \, .
\end{equation}
Several factors go into determining $\gamma(f)$ for space based systems. They include
the relative orientation and location of the detectors and the length of the
interferometer arms. The next section is devoted to calculating the overlap reduction
function for pairs of space based interferometers, and identifying which
configurations give the largest $\gamma(f)$, and hence the greatest sensitivity.

In analogy with our treatment of a single interferometer, we can define the
integrated signal-to-noise ratio:
\begin{equation}
{\rm SNR_{1\times 2}} = \frac{\vert \langle  s_1(t)s_2(t)\rangle\vert}
{\langle n^2_1(t) \rangle^{1/2}
\langle n^2_2(t) \rangle^{1/2}}
\end{equation}
and the signal-to-noise ratio at frequency $f$:
\begin{equation}
{\rm SNR}_{1\times 2}(f) = \frac{S_h(f) \vert {\cal R}_{12}(f)\vert}
{\sqrt{S_{n1}(f)S_{n2}(f)}} \, .
\end{equation}

We can improve upon these signal-to-noise ratios by
optimally filtering the cross-correlated signals. Suppose the detector outputs are
integrated over an observation time $T$:
\begin{equation}
    C(t) = \int_{t-T/2}^{t+T/2} dt' \int_{t-T/2}^{t+T/2} dt'' \, S_1(t')S_2(t'')
    Q(t'-t'') \, ,
\end{equation}
where $Q(t'-t'')$ is a filter function. The filter function is chosen
to maximize the integrated signal-to-noise ratio
\begin{equation}\label{snrc}
{\rm SNR}_C^2 = \frac{\langle C\rangle^2 }
    {\langle C^2 \rangle-\langle C \rangle^2} \, .
\end{equation}
The signal has expectation value
\begin{equation}\label{sigx}
    \langle C\rangle  = \frac{T}{5} \sin^2\beta\, \int^{\infty}_{-\infty} df
    \, S_h(f) \gamma(f) \widetilde{Q}(f)  \, ,
\end{equation}
and variance
\begin{equation}
\langle C^2 \rangle-\langle C \rangle^2
=\frac{T}{4}\int_{-\infty}^{\infty} \vert\widetilde{Q}(f)\vert^2
 M(f)\, df\, ,
\end{equation}
where
\begin{eqnarray}
M(f)&=&S_{n1}(f)S_{n2}(f)\left(1+{\rm SNR}_1(f)+{\rm SNR}_2(f) \right. \nonumber \\
&+& \left. {\rm SNR}_1(f){\rm SNR}_2(f)+{\rm SNR}^2_{1\times 2}(f)\right) .
\end{eqnarray}
In the limit that the signal-to-noise ratios are large, the variance is dominated
by the variance in the gravitational wave signal (cosmic variance):
\begin{equation}
M(f) \simeq S_h^2(f)\left({\cal R}_1(f){\cal R}_2(f)+{\cal R}_{12}^2(f) \right) \, .
\end{equation}

In general, the signal-to-noise ratio ${\rm SNR}_C$ will be a maximum for the
optimal filter\cite{paper1}
\begin{equation}
\widetilde{Q}(f)= \frac{S_h(f) \gamma^*(f) }{M(f) } \, .
\end{equation}
With this filter we have the optimal signal-to-noise ratio
\begin{equation}
    {\rm SNR}_C^2 = \frac{8 T}{25} \sin^4\beta \int_0^\infty df 
    \frac{\vert\gamma(f)\vert^2 S_h^2(f)}{M(f)} \, .
\end{equation}
The contribution to ${\rm SNR}_C$ from a frequency band of width $\Delta f$, centered
at $f$ is given by
\begin{eqnarray}
&& {\rm SNR}_C(f) \simeq \sqrt{2T\Delta f}\, {\rm SNR}_{1\times 2}(f)\left(1
+{\rm SNR}_1(f)\right. \nonumber \\
&&\hspace*{-0.02in}\left. 
+ {\rm SNR}_2(f) + {\rm SNR}_1(f){\rm SNR}_2(f)+{\rm SNR}^2_{1\times 2}(f)
\right)^{-\frac{1}{2}} \! .
\end{eqnarray}
The above approximation requires
\begin{equation}\label{cond}
\frac{\Delta f}{f} \ll \left(\frac{\partial \ln {\rm SNR}_i(f)}{\partial \ln f}\right)^{-1} \, .
\end{equation}
In the limit that the noise dominates
the signal we have
\begin{equation}\label{ndom}
{\rm SNR}_C(f) \approx \sqrt{2T\Delta f}\, {\rm SNR}_{1\times 2}(f) \, ,
\end{equation}
while in the limit that the signal dominates the noise we have
\begin{equation}
{\rm SNR}_C(f) \approx \sqrt{T\Delta f} \, .
\end{equation}
It is tempting to use (\ref{ndom}) to define an effective strain noise for the
cross-correlated system.
The difficulty with this approach is that at high frequencies $\gamma(f)$
oscillates rapidly and invalidates
the approximation (\ref{cond}) used to derive (\ref{ndom}). A better approximation
results from taking the sliding average
\begin{eqnarray}\label{snhigh}
{\rm SNR}^2_C(f) &\simeq & 2 T S_h^2(f) \int_{f-\Delta f/2}^{f+\Delta f/2} df' \frac{\vert
{\cal R}_{12}(f')\vert^2}{S_{n1}(f')S_{n2}(f')} \, , \nonumber \\
&& \nonumber \\
&\simeq & 2T\Delta f  S_h^2(f) \overline{\left(\frac{\vert
{\cal R}_{12}(f)\vert^2}{S_{n1}(f)S_{n2}(f)}\right)} \, .
\end{eqnarray}
Here the overbar denotes an average over the frequency interval $(f-\Delta f/2, f+\Delta f/2)$.
Using (\ref{snhigh}) we can define the effective sensitivity of the cross-correlated
detectors:
\begin{equation}\label{h12eff}
\tilde{h}_{{\rm eff}}(f)=\frac{1}{(2T\Delta f)^{1/4}} \overline{\left(\frac{\vert
{\cal R}_{12}(f)\vert^2}{S_{n1}(f)S_{n2}(f)}\right)}^{\; -1/4} \, .
\end{equation}
Unlike the corresponding expression (\ref{hef}) for a single detector, the effective
noise in a pair of cross-correlated detectors depends on the observation time $T$ and
the frequency resolution $\Delta f$. It is natural to choose a fixed frequency resolution
in $\ln f$, so that $\Delta f = \varepsilon f$. 

\begin{figure}[ht]
\vspace{55mm}
\includegraphics{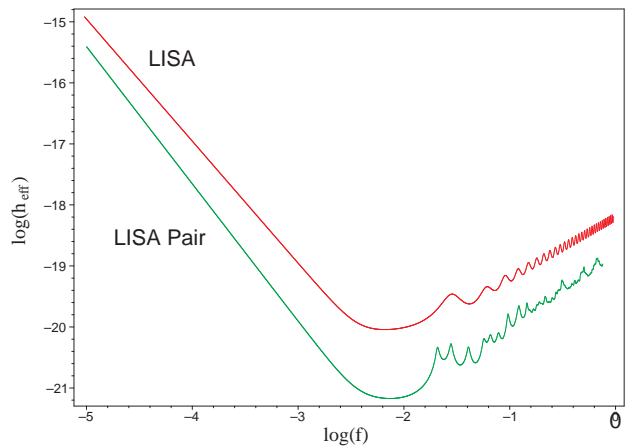}
\vspace{8mm}
\caption{The sensitivity of a single LISA interferometer compared to the sensitivity
of the optimally cross-correlated pair of LISA interferometers described in Section V.
The cross-correlation is for one year, with a frequency resolution of $\Delta f = f/10$.}
\end{figure}

Figure 3 compares the effective strain sensitivity of a pair of optimally cross-correlated LISA detectors
to the sensitivity of a lone LISA detector. The cross-correlated pair is $\sim 100$ times more
sensitive than a single detector across the frequency range 1 $\rightarrow$ 20 mHz. The sensitivity curve
for the cross-correlated interferometers scales as $f^{-9/4}$ for $f\ll f_*$ and $f^{5/4}$ for $f\gg f_*$.
This leads to a sharper ``V'' shaped sensitivity curve compared to a single interferometer where the
scaling goes as $f^{-2}$ for $f\ll f_*$ and $f$ for $f\gg f_*$.

\section{Optimizing the cross-correlation of two LISA detectors}

\begin{figure}[ht]
\vspace{50mm}
\includegraphics{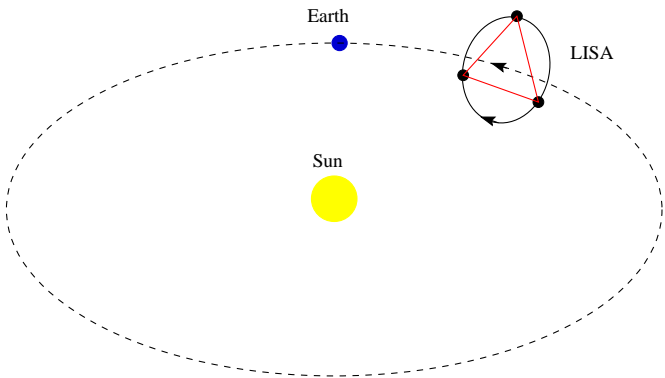}
\vspace{8mm}
\caption{The cartwheeling orbit of the LISA constellation. The dotted line is the
guiding center orbit and the solid line is the relative orbit of the
three spacecraft about the guiding center.}
\end{figure}

The LISA proposal\cite{lppa} calls for three identical spacecraft to fly in
an Earth trailing constellation at a mean distance from the Sun of 1 AU.
The spacecraft will maintain an almost constant separation of $L=5\times 10^9$ meters,
in a triangular configuration whose plane is inclined at $\pi/3$ radians to
the ecliptic. This is accomplished by placing each of the spacecraft on a
slightly inclined and eccentric orbit with a carefully chosen set
of initial conditions. The easiest way to derive the orbital parameters is to
start with all three spacecraft on a circular orbit with radius $R=1$ AU (the so-called
guiding center orbit), then introduce a small eccentricity and inclination to
each orbit. There is a unique configuration that keeps the distance between
all three spacecraft constant to leading order in the eccentricity $e$
(similar solutions exist for $N$ spacecraft). The orbits are inclined by
$i\simeq \sqrt{3} e$, and the constellation appears to rotate about the guiding center
on a circle with an inclination of $\pi/3$ and radius $2Re$. The relative rotation
of the constellation has the same period as the guiding center orbit.
The three spacecraft are evenly space about the circle a distance
$L\simeq 2\sqrt{3} Re$ apart (to leading order in the eccentricity).
The eccentricity is chosen to equal $e = 0.00965$ so that $L=5\times 10^9$ meters.
In a compromise between orbital
perturbations and communications costs, the plan is
to fly the constellation in an orbit that trails the Earth by 20 degrees.

\begin{figure}[ht]
\vspace{70mm}
\includegraphics{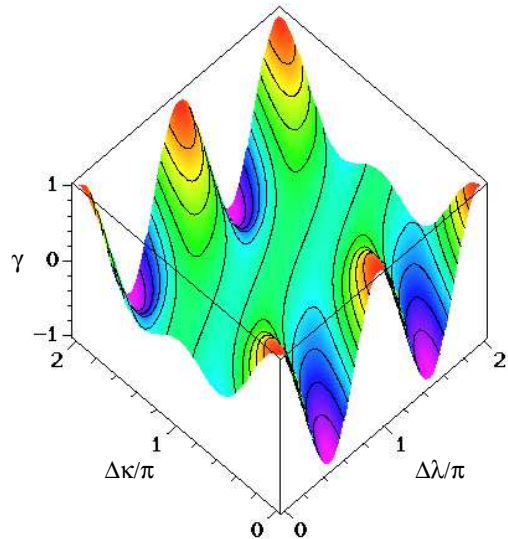}
\vspace{8mm}
\caption{The low frequency limit of orbit-averaged overlap reduction function, $\gamma(0)$,
as a function of $\Delta\kappa$ and $\Delta \lambda$.}
\end{figure}

It is natural to use an ecliptic coordinate system with the Sun at the origin to
describe the location of the LISA spacecraft. To leading order in $e$
the coordinates of each spacecraft are given by
\begin{eqnarray}\label{xyz}
x &=& a\cos(\alpha) + ae\left(\sin\alpha\cos\alpha\sin\beta 
-(1+\sin^2\alpha)\cos\beta\right)\nonumber \\
y &=& a\sin(\alpha) + ae\left(\sin\alpha\cos\alpha\cos\beta 
-(1+\cos^2\alpha)\sin\beta\right)\nonumber \\
z & = & \sqrt{3} a e \cos(\alpha-\beta) \, ,
\end{eqnarray}
where $a\simeq R$ is the semi-major axis, $\alpha = \omega t + \kappa$ is the phase of
the guiding center and $\beta=2n\pi/3+\lambda$ is
the relative phase of each spacecraft in the constellation $(n=0,1,2)$. If the
guiding center orbit does not lie in the plane of the ecliptic, we can obtain the
location of the spacecraft from (\ref{xyz}) by performing a rotation by an
angle $\iota$ about the axis $(\cos\xi,\sin\xi,0)$. The five constants $a$, $\kappa$,
$\lambda$, $\iota$ and $\xi$ fully specify a LISA constellation.

The cross-correlation of two LISA interferometers will depend on the relative orbits
of the two constellations. Unless the two interferometers share the same values of
$a$, $\iota$ and $\xi$, the distance between the corner spacecraft
in each interferometer, $d_{12}=\vert {\bf x}_1-{\bf x}_2\vert$,
will vary with time. The variation in $d_{12}$ translates into a variation of the
overlap reduction function, which poses a problem if we want to map the gravitational
wave background\cite{map}. Consequently, we shall set $\Delta a =\Delta \iota = \Delta\xi
=0$ and only consider constellations with different values of $\kappa$ and $\lambda$. 
When $\Delta \lambda =0$ we find the distance between corner spacecraft is given by
\begin{equation}
d_{12}=
\sqrt{2}a\, \sin\left(\frac{\Delta\kappa}{2}\right)\left(1-e\cos\alpha+\dots\right) \, .
\end{equation}
While this distance does vary with time, it is an order $e$ effect.
The situation is improved when $\Delta \kappa =0$ as the variation in $d_{12}$ drops
to order $e^2$:
\begin{equation}
d_{12} = 2\sqrt{2}ae\, \sin\left(\frac{\Delta\lambda}{2}\right)+{\cal O}(e^2) \, .
\end{equation}

\begin{figure}[ht]
\vspace{55mm}
\includegraphics{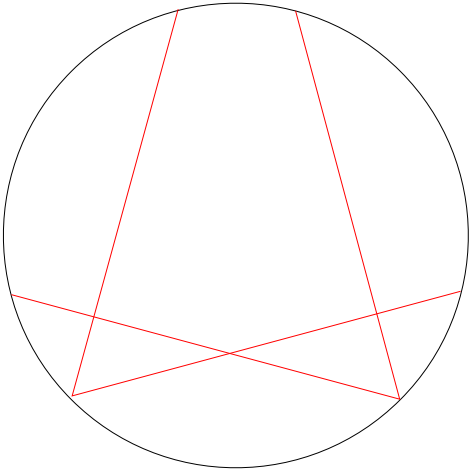}
\vspace{8mm}
\caption{The $\Delta\kappa=0$, $\Delta\lambda=\pi/2$ cross-correlation pattern.}
\end{figure}

There are two factors that go into determining the overlap reduction function $\gamma(f)$.
The first is the relative orientation of the arms in each interferometer, and the second
is the distance between the corner spacecraft. At low frequencies, the relative orientation of the
two interferometers is the dominant effect, while at high frequencies the distance between the
interferometers becomes important. Working in the zero frequency limit, the orbit-averaged
overlap reduction function is given by
\begin{eqnarray}\label{gamzero}
\gamma(0) &=& \frac{7}{64} - \frac{15}{32}\cos\Delta\kappa - \frac{41}{64}\cos^2\Delta\kappa
- \frac{7}{32}\cos\Delta\lambda \nonumber \\
&+& \frac{15}{16}\cos^2\Delta\lambda\cos\Delta\kappa
+ \frac{41}{32}\cos^2\Delta\lambda\cos^2\Delta\kappa \nonumber \\
&+& \frac{5}{16}\sin 2\Delta\lambda\sin 2\Delta\kappa
+\frac{3}{8}\sin 2\Delta\lambda\sin\Delta\kappa \, .
\end{eqnarray}
The magnitude of $\gamma(0)$ is maximized for $\Delta\kappa=0$ and $\Delta\lambda=0,\pi/2$
and $\pi$, as can be seen from the plot in Figure 5. Configurations with $\Delta\kappa=0$
are co-planar, and have the two interferometers phased by $\Delta\lambda$
about the small circle in Figure 4.
The $\Delta \lambda =0$ case is impractical as it places the two interferometers
on top of one another, but configurations with $\Delta \lambda \approx 0$ are a possibility. The
$\Delta \lambda = \pi/2$ configuration is shown in Figure 6. The
$\Delta \lambda = \pi$ case corresponds to the hexagonal cross-correlation studied by
Cornish \& Larson\cite{paper1}.

The distances $d_{12}(\Delta\kappa,\Delta\lambda)$ between the corner
spacecraft in each interferometer are:
\begin{eqnarray}
&& d_{12}(0,0)=0, \nonumber \\
&& d_{12}(0,\pi/2)=2\sqrt{2}\, ae = \sqrt{\frac{2}{3}}\, L\nonumber \\
&& d_{12}(0,\pi)=4ae = \frac{2}{\sqrt{3}}\, L \, .
\end{eqnarray}
As the frequency increases the
overlap reduction function decays due to the transfer functions ${\cal T}$ in the detector
response tensor, and from the overall factor of
$\exp(2\pi i f \widehat{\Omega}\cdot({\bf x}_1-{\bf x}_2))$ in (\ref{r12}):
\begin{eqnarray}
\gamma(f)_{0,0} = 1- \frac{169}{504}\left(\frac{f}{f_*}\right)^2
    + \frac{425}{9072}\left(\frac{f}{f_*}\right)^4 -\dots \nonumber \\ \nonumber \\ 
\gamma(f)_{0,\pi/2} = -1 + \frac{23}{42}\left(\frac{f}{f_*}\right)^2
    - \frac{3211}{27216}\left(\frac{f}{f_*}\right)^4 -\dots \nonumber \\ \nonumber \\ 
\gamma(f)_{0,\pi} = 1- \frac{383}{504}\left(\frac{f}{f_*}\right)^2+ 
    \frac{893}{3888}\left(\frac{f}{f_*}\right)^4 -\dots \nonumber \\
\end{eqnarray}
As expected, the magnitude of the overlap reduction function decays more rapidly for configurations with
larger values of $d_{12}$. On these grounds, the $\Delta\kappa=0$,
$\Delta \lambda \approx 0$ configuration would appear to be the best option.
However, it is also the configuration
most likely to suffer from correlated noise in the two interferometers. Taking all these factors into
account, we believe that the $\Delta\kappa=0$, $\Delta\lambda=\pi/2$ configuration represents
the optimal cross-correlation pattern.

\begin{figure}[ht]
\vspace{52mm}
\includegraphics{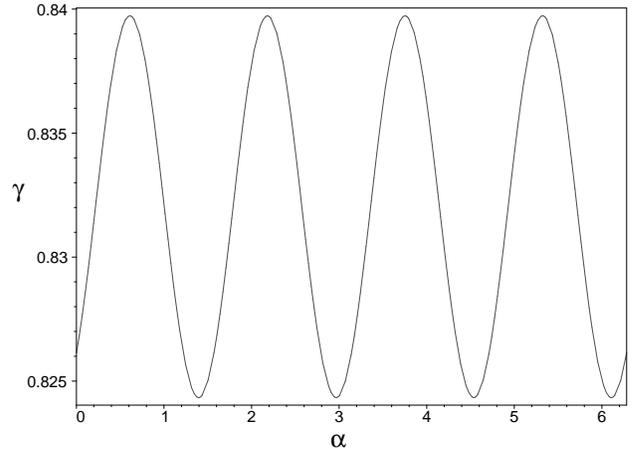}
\vspace{8mm}
\caption{Fluctuations in the zero-frequency overlap reduction function, $\gamma(0)$ over the course
of one orbit. The detector pair has $\Delta \kappa = 40^o$ and $\Delta \lambda = 20^o$.}
\end{figure}

\begin{figure}[ht]
\vspace{52mm}
\includegraphics{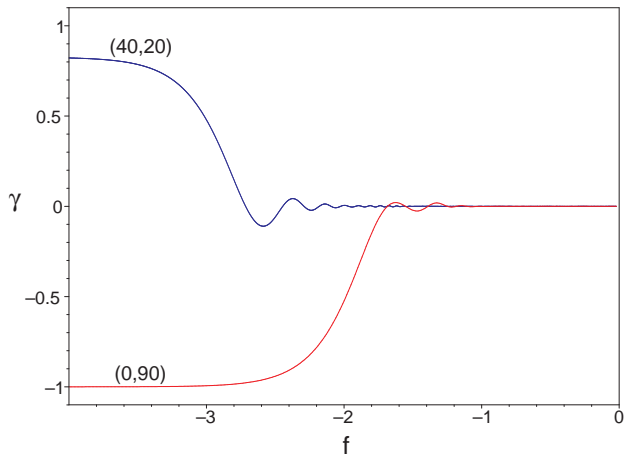}
\vspace{8mm}
\caption{The overlap reduction function $\gamma(f)$ for the $\Delta \kappa=40^o$,
$\Delta \lambda=20^o$ and $\Delta \kappa=0$, $\Delta \lambda=90^o$ cross-correlations.}
\end{figure}

Other factors may play a role in deciding how to deploy a pair of LISA detectors.
For example, if the priority is to determine the location of bright black hole binaries for
comparisons with X-ray observations, then it is advantageous to place the detectors far
apart. When the detectors are placed far apart, the phase of the waves arriving at the
two detectors gives directional information that compliments the usual amplitude and phase
modulation\cite{cc,hm}. Fixing a particular value for $\Delta\kappa$, we can optimize the cross-correlation
by maximizing $\vert \gamma(0) \vert$ according to equation (\ref{gamzero}). The full solution
is complicated, but a good approximation is to set $\Delta \lambda = \Delta \kappa/2$. 
For example, a second LISA constellation could be flown in an orbit that leads the Earth by $20^o$.
The angle between the leading and following detectors is then $\Delta \kappa = 40^o$.
As shown in Figure 7, the zero-frequency overlap reduction function
for this configuration fluctuates by $\sim1.5\%$ about a mean value of $\gamma(0)=0.833$.

The main disadvantage to having the interferometers separated by $\Delta \kappa = 40^o$ is that
the overlap reduction function decays rapidly above 1 mHz. The contrast between the 40 degree
option and the optimal cross-correlation is apparent in Figure 8. The sensitivity of a pair of
LISA detectors with $\Delta \kappa \neq 0$ and $\Delta \lambda =0$ was studied by
Ungarelli \& Vecchio\cite{uv}. Our conclusions differ from theirs as they neglected to include
the transfer functions ${\cal T}$ in the calculation of the overlap reduction function.
Moreover, the orbital parameters they used are not optimal.

\section{Detecting gravitational wave backgrounds}

We are now in a position to apply the results of the previous sections. As an illustration
we will consider two types of gravitational wave backgrounds: a cosmological gravitational wave
background (CGB) with a scale-invariant spectrum; and an astrophysical background produced by
galactic and extra-galactic White Dwarf binaries. Plots of the one-sided power spectral densities
for these sources are shown in Figure 9, along with the projected noise in each interferometer.
The CGB power spectrum is for a scale invariant inflationary model with an energy density per
logarithmic frequency interval of $\Omega_{\rm gw}(f)=10^{-14}$. This quantity
is related to the power spectral density by
\begin{equation}
    S_h(f) = \frac{3H_0^2}{4 \pi^2} \frac{\Omega_{\rm gw}(f)}{f^3}  \, ,
\end{equation}
where $H_0\simeq 65$ km s$^{-1}$ Mpc$^{-1}$ is the Hubble constant.
The White Dwarf power spectrum is
taken from the work of Bender \& Hils\cite{bhil}, and the noise power spectrum is estimated from the
noise budget in the LISA Pre-Phase A report\cite{lppa}.

\begin{figure}[ht]
\vspace{52mm}
\includegraphics{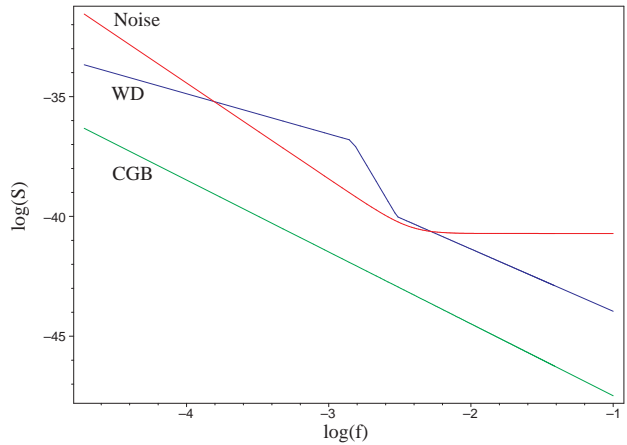}
\vspace{8mm}
\caption{One-sided power spectral densities, $S_h(f)$, for the CGB and the confusion
limited White Dwarf background. The anticipated noise spectral density, $S_n(f)$, for LISA is also
shown.}
\end{figure}

Using these power spectra we can calculate the optimal filters for detecting each background with the
optimally cross-correlated LISA interferometers. The White Dwarf filter is
shown in Figure 10 and the CGB filter is shown in Figure 11.
We see from these plots that the bulk of the cross correlation occurs for signals that
are lagged by less than the light travel time in the interferometer, $2L\sim 30$ seconds.
In the frequency domain, the bulk of the cross-correlation occurs across the floor region
(1 and 20 mHz) of the LISA sensitivity curve. The White Dwarf filter favours slightly higher
frequencies than the CGB filter due to the peak in the White Dwarf spectrum at 2 mHz.

\begin{figure}[ht]
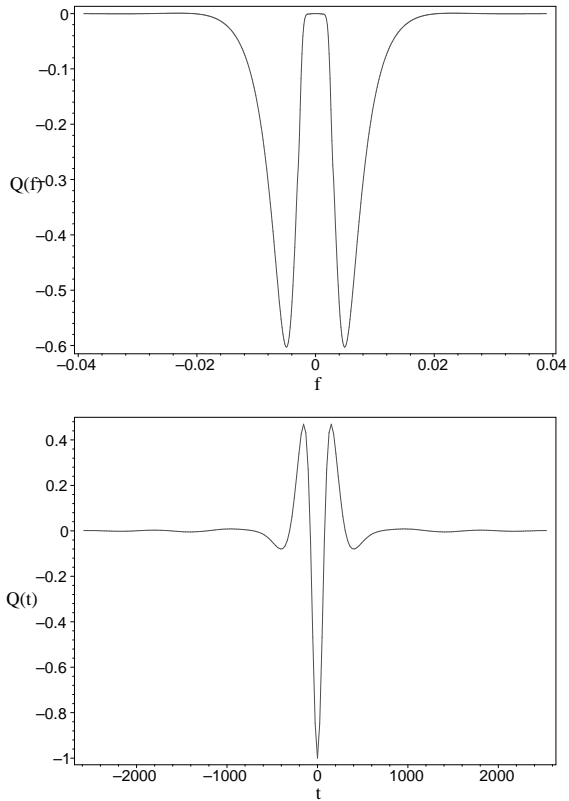

\vspace{100mm}
\includegraphics{QfWD1.ps}
\includegraphics{QtWD1.ps}
\vspace{8mm}
\caption{The optimal filter for detecting the White Dwarf background with a pair of
LISA detectors. The upper panel is in the
frequency domain (Hertz) and the lower panel is in the time domain (seconds).}
\end{figure}

\begin{figure}[ht]
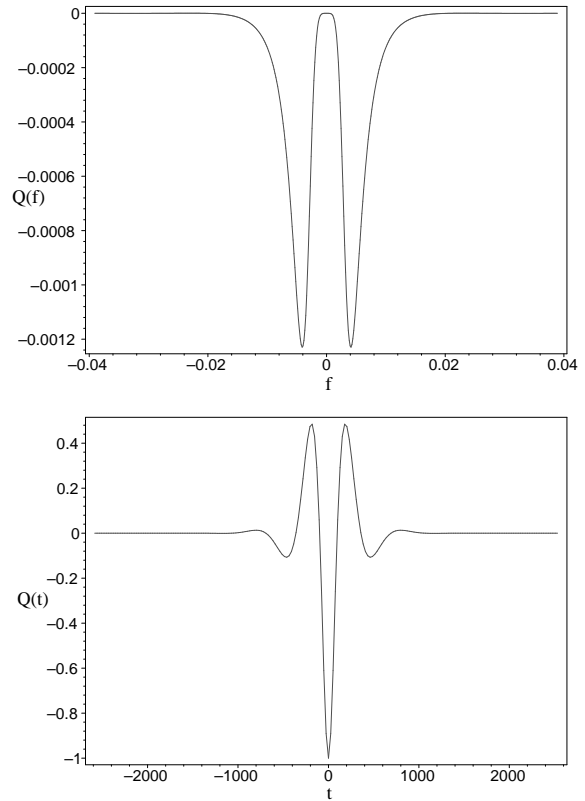

\vspace{100mm}
\includegraphics{QfCGB1.ps}
\includegraphics{QtCGB1.ps}
\vspace{8mm}
\caption{The optimal filter for detecting the CGB with a pair of LISA detectors. The upper panel is in
frequency domain (Hertz) and the lower panel is in the time domain (seconds).}
\end{figure}

Using the filters shown in Figures 10 and 11, the LISA
pair could detect the sources described in Figure 9 with an integrated signal-to-noise ratio of
${\rm SNR}=0.07$ for the CGB and ${\rm SNR}=86.2$ for the White Dwarf binaries. These numbers
are calculated from (\ref{snrc}) using $T=1$ year.

To detect a stochastic background with 95\% confidence requires a signal-to-noise
ratio of ${\rm SNR}=2$\cite{bruce}. By rescaling the Bender-Hils estimate\cite{bhil}
for the White Dwarf power spectrum, and taking into account the changes this makes in
the shape of the optimal filter, we find that the LISA pair could still detect the
White Dwarf background even if the spectral density were 1200 times lower than the level
shown in Figure 9. Alternatively, the White Dwarf background shown in Figure 9 can be
detected with greater than 95\% confidence after just five hours of observations. The
prospects are not so promising for the cosmological background, as the
CGB would have to have an energy density twenty-eight times larger
than the level shown in Figure 9 to be detectable after one year.
This exceeds existing limits\cite{lk} on the gravitational
wave energy density in scale-invariant inflationary models by a factor of $\sim 15$, but other
more exotic models may produce a detectable signal.

\section*{Acknowledgements}

I would like to thank Peter Bender and Bill Hiscock for their input concerning the
Sagnac interferometer and the White Dwarf background. I am indebted to
Massimo Tinto for pointing out an error
in my original calculation of the noise spectral density for the Sagnac signals.
I benefitted from many lengthy discussion with Shane Larson. This work was supported by
NASA grant NCC5-579.

\section*{Appendix: Noise spectral density}

The various interferometer signals are built from phase measurements taken at
each spacecraft. The phase measurements record the phase
difference between the incoming and local laser signals. Taking a
simplified model of the LISA system with one laser on board each spacecraft, there
will be six such readouts. We label the phase measurement made at time $t$ by $\Phi_{ij}(t)$, where
the first index refers to the spacecraft that sends the signal, and the second index
refers to the spacecraft that receives the signal. The time-varying part of phase
has contributions from laser phase noise $C(t)$, gravitational wave strain $\psi(t)$,
shot noise $n^s(t)$, and acceleration noise ${\bf n}^a(t)$:
\begin{eqnarray}
&&\Phi_{ij}(t)=C_i(t-L_{ij})-C_j(t)+\psi_{ij}(t)+n^s_{ij}(t)\nonumber \\
&& \hspace*{0.8in} -\widehat{{\bf x}}_{ij}\cdot({\bf n}^a_{ij}(t)
-{\bf n}^a_{ji}(t-L_{ij})) \, .
\end{eqnarray}
Here $L_{ij}=L_{ji}$ is the distance between spacecraft $i$ and $j$, and $\widehat{{\bf x}}_{ij}$
is one of the three unit vectors defined in figure 1, {\it eg.} $\widehat{{\bf x}}_{12}
=-\widehat{{\bf x}}_{21}={\bf a}$. The gravitational wave strain is given by
\begin{equation}
\psi_{ij}(t) =
\frac{{\bf h}(f,t-L_{ij},{\bf x}_i):(\widehat{{\bf x}}_{ij}\otimes\widehat{{\bf x}}_{ij})
{\cal T}(\widehat{{\bf x}}_{ij}\cdot\widehat{\Omega},f)}{2L_{ij}} \, .
\end{equation}
The shot noise $n^s_{ij}(t)$ is from the photo-detector in spacecraft $j$ measuring
the laser signal from spacecraft $i$, while the acceleration noise ${\bf n}^a_{ij}(t)$
is due to the accelerometers in spacecraft $j$ that are mounted on the optical assembly
that points toward spacecraft $i$.

The basic Michelson signal extracted from vertex 1 has the form
\begin{eqnarray}\label{mnoise}
S_1(t)&=&\Phi_{12}(t-L_{12})+\Phi_{21}(t)-\Phi_{13}(t-L_{13})-\Phi_{31}(t) \nonumber \\
&=& s_1(t)+C_1(t-2L_{12})-C_1(t-2L_{13})\nonumber \\
&& +n^s_{12}(t-L_{12})++n^s_{21}(t) \nonumber \\
&& -n^s_{13}(t-L_{13})-n^s_{31}(t)\nonumber \\
&& -2 {\bf a}\cdot{\bf n}^a_{12}(t-L_{12})-2{\bf c}\cdot{\bf n}^a_{13}(t-L_{13})\nonumber \\
&& +{\bf a}\cdot({\bf n}^a_{21}(t)+{\bf n}^a_{21}(t-2L_{12}))\nonumber \\
&& +{\bf c}\cdot({\bf n}^a_{31}(t)+{\bf n}^a_{31}(t-2L_{13}))\, . 
\end{eqnarray}
The gravitational wave contribution, $s_1(t)$, is given by equation (\ref{m1}). The laser
phase noise from the corner spacecraft are automatically canceled, but the phase noise
from the vertex laser will dominate the response unless $L_{12}=L_{13}$ to high precision.
So long as $L_{12}\approx L_{13} \approx L$, the remaining phase noise can be eliminated
by differencing the Michelson signal with a copy from time $2L$ earlier\cite{ta}. For
simplicity we will set $L_{12}=L_{13}=L$ in (\ref{mnoise}) to estimate the noise spectral
density $S_n(f)=\langle n(f) n^*(f) \rangle$:
\begin{eqnarray}
S_n(f)&=&S^s_{12}(f)+S^s_{21}(f)+S^s_{13}(f)+S^s_{31}(f) \nonumber \\
&& 4\cos^2(f/f_*)\left( S^a_{21}(f)+S^a_{31}(f)\right)
\nonumber \\
&& +4S^a_{12}(f) +4S^a_{13}(f)\, .
\end{eqnarray}
Assuming that each detector has the same noise spectral density we have
\begin{equation}
S_n(f) = 4 S_s(f)+8(1+\cos^2(f/f_*))S_a(f) \, .
\end{equation}
The $\cos^2(f/f_*)$ term comes from combining the acceleration noise in spacecraft 1
at times $t$ and $t-2L$.
The LISA pre Phase A report\cite{lppa} quotes the shot noise in terms of the power
spectral density of optical-path length fluctuations over a path of length $L=5\times 10^9$ m:
\begin{equation}
S_{\rm shot} = 1.21 \times 10^{-22} \quad {\rm m}^2 \; {\rm Hz}^{-1} \, .
\end{equation}
This can be converted to strain spectral density by dividing by the path length squared:
$S_s(f) = S_{\rm shot}/L^2 = 4.84 \times 10^{-42} \; {\rm Hz}^{-1}$. Each inertial sensor
is expected to contribute an acceleration noise with spectral density
\begin{equation}
S_{\rm accl} = 9\times 10^{-30} \quad {\rm m}^2 \; {\rm s}^{-4}\; {\rm Hz}^{-1} \, .
\end{equation}
To convert this into phase noise we need to divide by path length squared, and by
the angular frequency of the gravitational wave to the fourth power:
\begin{equation}
S_a(f) = 2.31 \times 10^{-40} \left(\frac{{\rm mHz}}{f}\right)^4 \quad {\rm Hz}^{-1} \, .
\end{equation}
Thus,
\begin{eqnarray}
S_n(f)&=&1.85\times 10^{-39}\left(\frac{{\rm mHz}}{f}\right)^4
\left(1+\cos^2\left(\frac{f}{9.55 {\rm mHz}}\right)\right) \nonumber \\
&& + 1.94\times 10^{-41} \nonumber \\
&\simeq& 3.7 \times 10^{-39}\left(\frac{{\rm mHz}}{f}\right)^4
+ 1.94\times 10^{-41} \, .
\end{eqnarray}
This result differs slightly from the noise calculation given in Ref.\cite{paper1}.
The factor of four difference at low frequencies
can be traced to our dividing by $(2L)^2$ rather than $L^2$
in the conversion from position to strain noise spectral density in the earlier
calculation.

The Sagnac signal extracted at vertex 1 is given by
\begin{eqnarray}
S_1(t)&=&\Phi_{13}(t-L_{23}-L_{12})+\Phi_{32}(t-L_{12})+\Phi_{21}(t)\nonumber \\
&& -\Phi_{12}(t-L_{23}-L_{13})-\Phi_{23}(t-L_{13})-\Phi_{31}(t)\, .\nonumber \\
\end{eqnarray}
Laser phase noise cancels exactly in the Sagnac signal for any arm lengths $L_{ij}$.
Specializing to the
case where all the arm lengths are approximately equal and each optical assembly has the same
noise spectrum, the remaining noise sources
combine to give a noise spectral density of
\begin{equation}
S_n(f)=6S_s(f)+8\left(\sin^2(3f/2f_*)+2\sin^2(f/2f_*)\right)S_a(f) \, .
\end{equation}
The symmetrized Sagnac signal is given by
\begin{eqnarray}
S(t)&=&\frac{1}{3}\left( \Phi_{21}(t)+\Phi_{32}(t-L_{12})+\Phi_{13}(t-L_{23}-L_{12})
\right. \nonumber \\
&& -\Phi_{31}(t)-\Phi_{23}(t-L_{13})-\Phi_{12}(t-L_{23}-L_{13}) \nonumber \\
&& +\Phi_{23}(t)+\Phi_{13}(t-L_{23})+\Phi_{21}(t-L_{23}-L_{13}) \nonumber \\
&& -\Phi_{12}(t)-\Phi_{31}(t-L_{12})-\Phi_{23}(t-L_{12}-L_{13}) \nonumber \\
&& +\Phi_{13}(t)+\Phi_{21}(t-L_{13})+\Phi_{32}(t-L_{13}-L_{12}) \nonumber \\
&& \left. -\Phi_{23}(t)-\Phi_{12}(t-L_{23})-\Phi_{31}(t-L_{23}-L_{12})
\right),\nonumber \\
\end{eqnarray}
from which it follows that the noise spectral density equals
\begin{eqnarray}
S_n(f)&=& \frac{2}{3}\left(1+2\cos(f/f_*)\right)^2\left(S_s(f)\right. \nonumber \\
&& \left. \quad + 4\sin^2(f/2f_*)S_a(f)\right) \, .
\end{eqnarray}
The overall factor of $(1+2\cos(f/f_*))^2$ cancels the corresponding factor that appears
in the signal spectral density $S_h(f)$ for the symmetrized Sagnac interferometer.


\begin{thebibliography}{99}
\bibitem{foot} To ensure that the Sagnac and
Michelson signals have the same number of noise contributions from each arm, the output of the
three Michelson interferometers are averaged and compared to two-thirds of the symmetrized Sagnac
signal.
\end{thebibliography}
\end{document}